# To Centralize or Not To Centralize: A Tale of Swarm Coordination

Justin Hu, Cornell University
Ariana Bruno, Cornell University
Drew Zagieboylo, Cornell University
Mark Zhao, Cornell University
Brian Ritchken, Cornell University
Brendon Jackson, Cornell University
Joo Yeon Chae, Cornell University
Francois Mertil, Cornell University
Mateo Espinosa, Cornell University
Christina Delimitrou, Cornell University

{jh2625,amb633,dz333,yz424,bjr96,btj28,jc2464,fjm83,me326,delimitrou}@cornell.edu

## Abstract

Large swarms of autonomous devices are increasing in size and importance. When it comes to controlling the devices of large-scale swarms there are two main lines of thought. *Centralized control*, where all decisions - and often compute - happen in a centralized back-end cloud system, and *distributed control*, where edge devices are responsible for selecting and executing tasks with minimal or zero help from a centralized entity. In this work we aim to quantify the trade-offs between the two approaches with respect to task assignment quality, latency, and reliability. We do so first on a local swarm of 12 programmable drones with a 10-server cluster as the backend cloud, and then using a validated simulator to study the tail at scale effects of swarm coordination control. We conclude that although centralized control almost always outperforms distributed in the quality of its decisions, it faces significant scalability limitations, and we provide a list of system challenges that need to be addressed for centralized control to scale.

## 1 Introduction

Swarms of autonomous edge devices are increasing in number, size, and popularity [4, 7]. From UAVs, and blips, to self-driving cars, and supply-chain robots, swarms are enabling new distributed applications.

There are two dimensions of execution in swarms that present interesting system tradeoffs: *where application computation happens*, and *where coordination control or task assignment happens*. The two have often been viewed as a single problem, with control and application execution either happening in a centralized cloud, or distributed in a single or multiple edge devices. The trade-offs of remote versus local execution have been particularly well studied not only for swarms of autonomous devices, but traditional mobile and embedded devices as well [8, 9, 16, 20]. In this work we argue that the two problems, although connected, present each unique challenges and opportunities, and focus on the latter design question of where swarm coordination control happens.

Similar to cloud systems, the resource manager in swarms must guarantee performance, efficiency, and responsiveness. Unlike cloud systems though, coordination control in swarms has to additionally account for challenges including unreliable communication channels, limited battery capacity, and constrained on-board resources when assigning tasks to edge devices. Resource managers in the cloud follow several designs, the most popular being *centralized* [11–13, 21], *two-level* [17], and *distributed* [14, 15, 18, 19]. Centralized cluster managers are typically superior in terms of scheduling quality, since they have a global view of the cluster state and can make high quality resource assignment decisions. At the other extreme, distributed schedulers optimize for scheduling latency, often sacrificing some scheduling quality in the process. Distributed cluster schedulers come in different flavors, the two most prominent being *shared-state* designs, where each of the scheduling agents can allocate resources across the entire cluster [19], and designs where each scheduling agent operates on a partition of the cluster resources. The former must account for conflicting decisions between scheduling agents, while the latter sacrifices some decision quality for a simpler, more scalable design.

Coordination control in swarms has similar trade-offs. One line of research argues for centralized control where a single system, typically residing in a cloud system is responsible for task assignment, monitoring, and control of all edge devices [7]. There are several advantages to this approach. First, a centralized cloud controller *pushing* tasks to autonomous devices has virtually unlimited resources compared to the heavily resource-constrained edge devices, which improves its decision quality. Similarly a centralized controller can further improve the quality of its decisions by aggregating data from multiple devices, e.g., in applications like commute routing, or disaster recovery (we will ignore the security implications of cross-device data sharing in this work). Unfortunately centralized controllers also constitute bottlenecks as the size and heterogeneity of the swarm increases, and as tasks become shorter and more fine-grain as edge devices adopt serverless microservices.

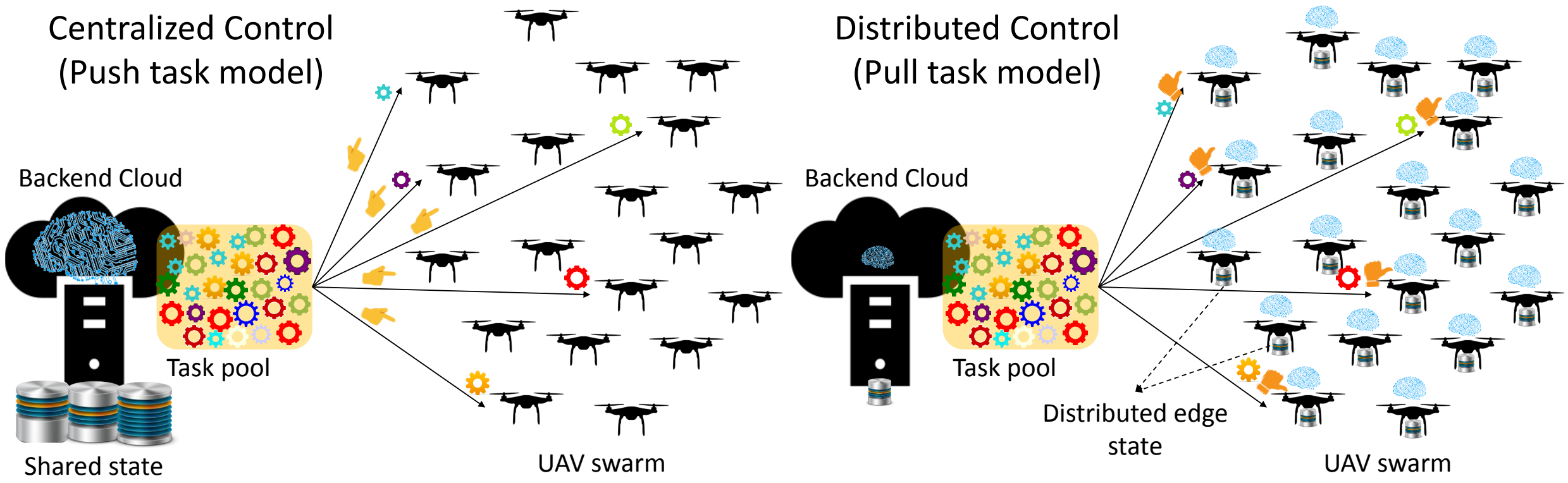

Figure 1: (a) Centralized coordination control using a back-end cloud that aggregates state across edge devices and has global visibility of the task pool and state of all devices, (b) Distributed control where each of the edge devices pulls tasks from the global task pool with only local knowledge of its state and resources.

On the other hand, distributed control where each drone independently *pulls* tasks to execute scales better [4], but sacrifices decision quality for decision latency, as edge devices lack a global view of the swarm state, and can only dedicate a small fraction of their limited resources to selecting suitable tasks to perform. Note that even in a distributed control framework the task execution can happen locally on the edge device, or the edge device can simply record sensor data and offload part of all of the computation to the back-end cloud.

The goal of this paper is to compare these two models of coordination control, and identify system bottlenecks for each. We first use a local swarm of 12 programmable Parrot AR2.0 drones [6] that perform routing, image recognition, and obstacle avoidance, and show that centralized control clearly outperforms the distributed control framework for small swarms. We then focus on the scalability of each approach through simulation of swarms of thousands to tens of thousands of autonomous devices. Centralized control is able to maintain its high quality decisions, even in the presence of device heterogeneity, and failure-prone communication channels, but at the price of prohibitively high scheduling latency. The more diverse and unreliable swarms become, the wider the differences between centralized and distributed control. We conclude with a what-if comparison between centralized and distributed coordination control, assuming bottlenecks including network processing latency, and scheduling serialization are alleviated, to highlight the system steps that need to be taken for centralized swarm control to realize its potential.

## 2 Centralized vs. Distributed Control

We examine two systems for coordination control. First, a centralized system where all control decisions happen in a back-end cloud and tasks are *pushed* to the edge devices. In this case the edge devices are perceived as "unintelligent" endpoints that collect sensor data and transfer it to the cloud, and run some local lightweight tasks. The centralized controller has global visibility on the state of the entire swarm, including each device's location, sensors, compute and memory resources, and fine-grain battery availability.

The alternative is a distributed coordination system, where each edge device *pulls* tasks to execute, having only local visibility in its own on-board resources, location, and battery. Fig 1 shows an overview of the two systems. We assume that the up-to-date pool of available tasks is persistently maintained in the back-end cloud. Hybrid scenarios where a subset of control decisions are made by edge devices, or where groups of edge devices are treated as an entity by a centralized controller are possible, however in this work we want to study the two systems in their extreme. Below we describe the design of the centralized and distributed swarm controllers.

**Centralized control design:** The centralized controller is a multi-threaded runtime implemented in C++, and runs in a dedicated 2-socket, 40-core server of the back-end cloud. While there are tasks in the global task pool, it assigns them in FIFO order. For each task, the controller first filters drones by their sensors, and discards those that do not have the sensors required for the task. Then it orders the filtered drones based on their location and battery. The controller has performance and power models trained on the drones of our local swarm (see Section 3) to estimate the task execution time, and the battery budget a task will require, including the battery depletion from a drone moving to the location where the task needs to be executed [8, 16, 20]. Once it selects the most suitable drone, it sends the task description (*mission*) to the edge device over TCP. [1] The policy the controller currently uses is designed to deplete the battery of the assigned drone as little as possible; alternative policies that optimize solely for proximity or load balancing are also possible. Once the task is assigned to the drone, the controller updates the status of the drone as `busy`, and does not assign another task to it until its previous mission is completed.

**Distributed control design:** Each drone runs a copy of the distributed controller, also implemented as a C++ runtime. If a drone does not have a task currently assigned to it, it examines the global task pool in FIFO order. The task selection process is similar to the centralized controller; specifically the drone selects the first task that requires sensors the drone has, and whose battery requirements are within the drone's battery budget, including any battery consumed to move to where the task should be executed. Each drone has the same performance and power models as the centralized controller above. Once a drone pulls a task, it sets itself as `busy`, and does not select another task until its current workload is completed. Given that each drone is pulling tasks from the same global task pool conflicts are possible. For now we use *optimistic lock-free concurrency* for conflict resolution, a technique that has previously been applied to cluster schedulers [19]. Under optimistic concurrency if two or more drones select the same task, the first one succeeds (based on the global order in the persistent copy of the task pool), and the rest fail, and retry. Although conflicts increase scheduling latency, they are rare in practice, especially in heterogeneous swarms.

# 3 Evaluation

## 3.1 Methodology

**Small-scale swarm prototype:** We use 12 programmable Parrot AR 2.0 drones as our testbed. The centralized cloud is a cluster of 10 2-socket 40-core servers with 128GB of RAM each. All servers are connected to a 40Gbps ToR switch over 10Gbe NICs. Communication between the drones and cloud happens over TCP via a wireless router. All cloud servers are within 5 meters from the router, while the drones can move up to 50 meters of radius and 20 meters of altitude from the router. We are experimenting with alternative network configurations that allow better mobility as part of future work.

The applications the drones execute include routing, different versions of image recognition for people, buildings, trees, and other drones, and obstacle avoidance. The face, building, and tree image recognition are using a node.js `OpenCV` module, while the drone detection application is embedded in the `Parrot AR-Drone SDK` in node.js. The drone routing, motion control, and obstacle avoidance are done through the `ardrone-autonomy` library [1]. Finally, we use `cylon` [3] to send group commands, e.g., routing, to multiple edge devices over the router at once. Any of the applications can run on the edge devices, or in the back-end cloud. Execution in the cloud happens either natively, or over *Fission* [2], an open-source serverless framework. We are exploring the tradeoffs between edge, cloud native, and serverless execution as part of ongoing work.

We also developed an end-to-end tracing system using OpenTracing [5] for monitoring scheduling and computation on the cloud. Monitoring on the drones is done with wireshark for network requests, and a logger embedded in the Parrot SDK for computation. The tracing framework has no meaningful impact on performance (less than 0.1% on tail latency (scheduling and task execution), and 0.2% on task scheduling throughput).

**Large-scale swarm simulation:** To study the tail at scale effects of swarm coordination beyond what our local prototype allows, we have also developed a scalable, event-driven swarm simulator in Python. The simulator models data transfer latencies, compute and network contention, battery depletion at the edge, and UAV heterogeneity and unreliability where applicable. We have validated the simulator against our drone swarm, and showed it achieves less than 2% deviation in scheduling latency, and less than 5% in task execution time. We are further refining the simulator to capture memory contention on the cloud and edge devices. By default we simulate `1,000` homogeneous Parrot AR-2.0drones, and introduce different swarm sizes, heterogeneity, and communication unreliability in latter experiments.

Unless otherwise specified, the workload unless is a uniform mix of routing, and people, building, or drone recognition tasks. Recognition tasks come with a conditional obstacle avoidance task in the event where an obstacle is detected. There are always enough tasks in the task pool that neither the centralized scheduler nor the drones are starved.

## 3.2 Coordination Control Trade-offs

**Scheduling latency:** Fig. 2(a,c) show the scheduling latency Cumulative Distribution Functions (CDFs) in the centralized and distributed control frameworks, for the two swarms (small- and large-scale). In the real prototype, the difference in scheduling speed is marginal, as the back-end cluster can easily keep up with the assignment of tasks. The difference becomes substantial for the large-scale swarm. While the distributed control framework is able to maintain its scheduling latency almost unchanged, the centralized framework introduces scheduling latencies comparable, and in some cases longer than the task duration. This is not surprising given that the centralized framework attempts to find the most suitable drone when allocating a task.

[1] The controller maintains open sockets with each of the drones to avoid long instantiation overheads.

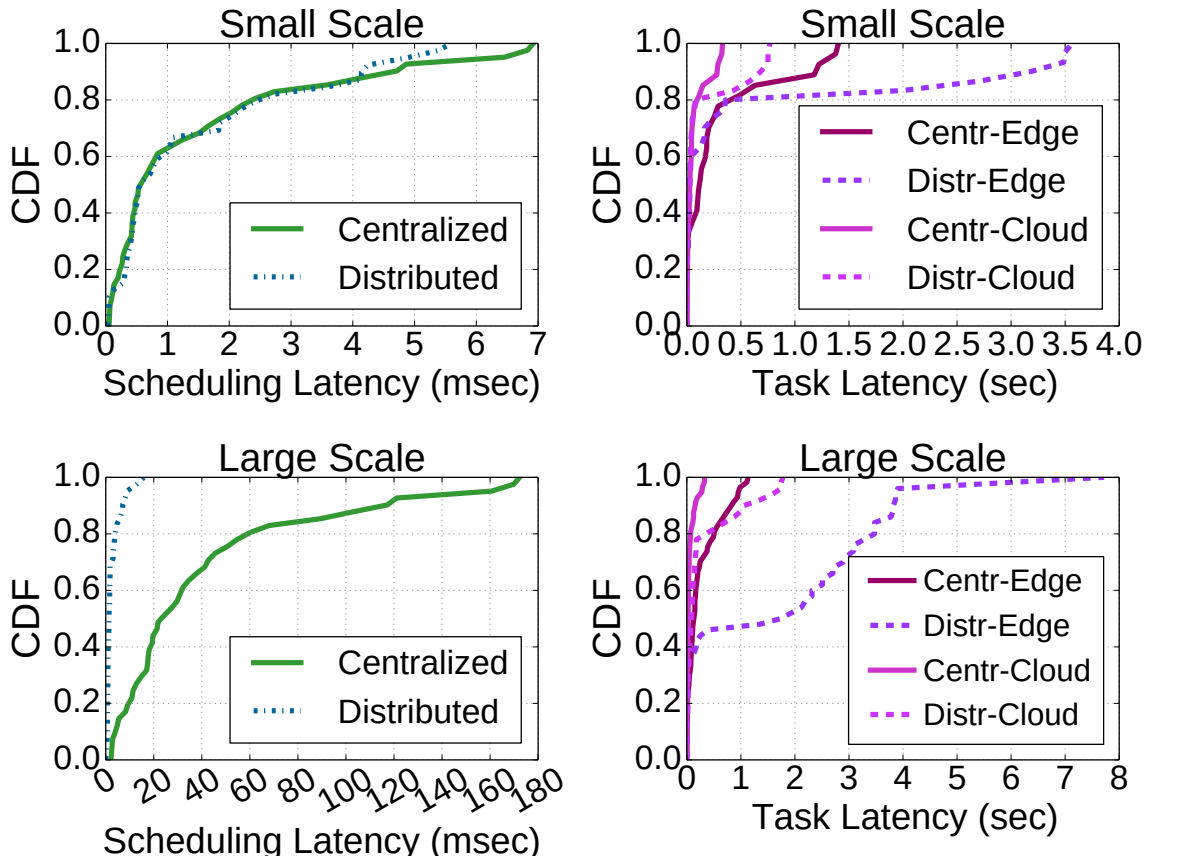

Figure 2: Comparison of scheduling speed and scheduling quality between centralized and distributed control frameworks. The top figures show results from our local swarm prototype, and the bottom figures simulation results across a 1000 UAV swarm.

**Scheduling quality:** Fig. 2(b,d) show the opposite tradeoff. When employing centralized coordination control the majority of tasks achieve low latency; this is the case regardless of whether computation happens locally at the edge, or in the back-end cloud. Applications in the cloud run in regular Docker containers; porting the same applications over the Fission serverless framework increased average latency by 6%, primarily due to Fission's request handling overhead. In general task execution time is longer at the edge due to the limited on-board resources on the drones, although tasks like routing and obstacle avoidance perform almost the same on the drones and back-end cloud. The difference between centralized and distributed control in this case is due to the centralized controller having a global view of the swarm and tasks, and making higher quality scheduling decisions. On the opposite side, each individual drone only has visibility of its own state and resources, leading to suboptimal task assignment decisions. Neighboring drones exchanging information about their availability and resources could help bridge the gap between centralized and distributed control quality.

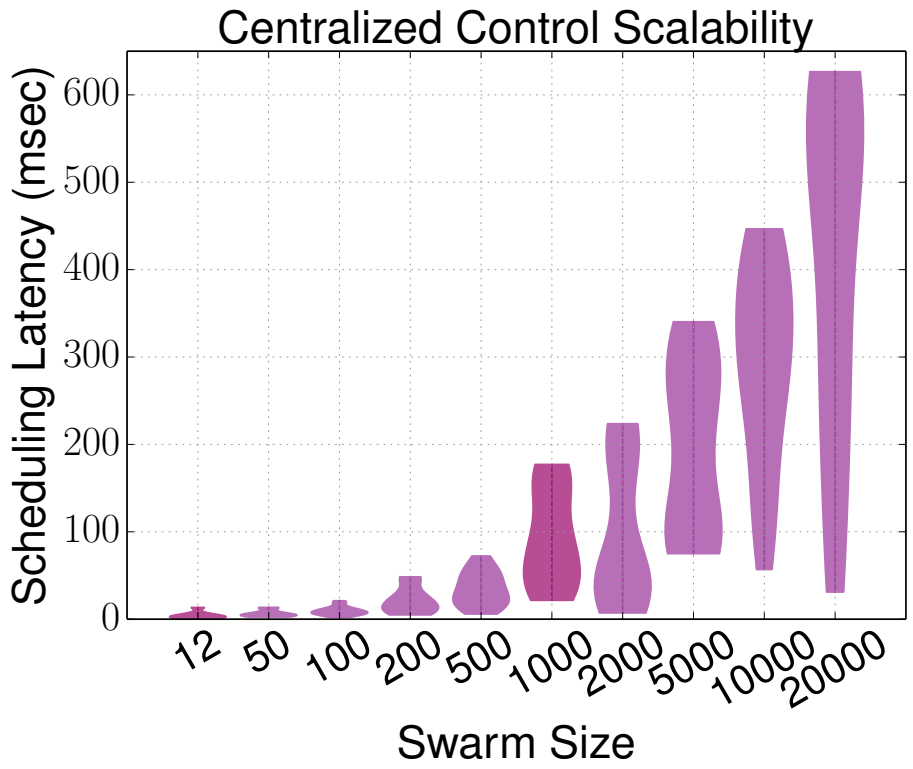

Figure 3: Violin plots of scheduling latency for the centralized control framework as the size of the swarm increases. The leftmost violin plot is obtained from our local swarm prototype; the other violin plots are obtained through simulation.

**Control scalability:** So far the large-scale swarm was configured to 1,000 homogeneous drones. We now examine the scalability of centralized control as the size of the swarm scales. Fig. 3 shows violin plots of scheduling latency distribution for each swarm size. The leftmost violin plot is obtained on the physical prototype, while the other results are obtained through simulation. The figure shows two effects. First, as expected, as the size of the swarm increases the magnitude of scheduling latency for the centralized framework increases rapidly. Although latency is not quite linear with the size of the swarm, centralized control faces significant challenges in large swarms. More interestingly, the figure shows a change in the shape of the violin plot as swarms increase. For small swarms the majority of tasks have low scheduling latency, and only a few outliers take longer to assign. In the middle region of the figure, violin plots start having two peaks; one corresponding to tasks scheduled quickly, and one to tasks incurring significant overheads to schedule. The larger the swarm becomes the higher the peak corresponding to slow-scheduled tasks, with swarms of 10,000 agents or more incurring high scheduling latency to the majority of incoming tasks.

**Imperfect swarms:** Finally, we revisit the results above in the presence of edge device heterogeneity, and unreliable communication channels between the cloud and swarm. Fig. 4(a,b) show the scheduling latency for the small and large swarms (1,000 drones) when the drones are heterogeneous in their sensors, battery deposit, and CPU frequency. In the real prototype, we disable a subset of available sensors to introduce heterogeneity, and initiate the experiment with some drones partially charged. We also use DVFS in the on-board CPU to lower its frequency. While latency in the small-scale swarm is not significantly impacted, the centralized controller in-troduces considerably higher scheduling latencies in the large swarm in the presence of heterogeneity, as it takes longer to identify a device with suitable resources for a given task. Heterogeneity has a significant impact on task execution time as well. Task execution time with the centralized controller again outperforms the distributed framework, however the difference is now more pronounced as there are clear differences between each edge device's capabilities.

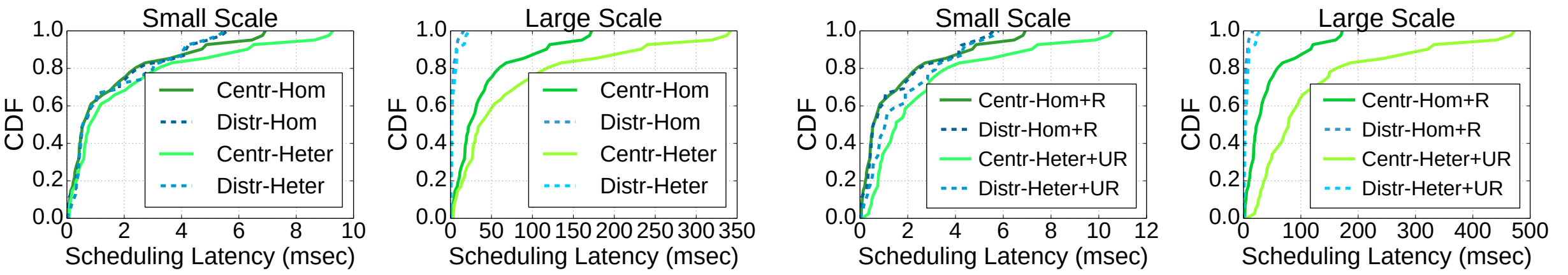

Figure 4: Comparison of scheduling speed in centralized and distributed control frameworks when (a,b) the swarm is heterogeneous, and when (c,d) the swarm is heterogeneous and unreliable (random failures).

Fig. 4(c,d) additionally introduces unreliable communication channels between cloud and edge devices. In the real prototype we randomly select 10% of drones to make unreachable over the network every few seconds. In the simulator, we similarly disconnect 10% of the drones over short intervals. Scheduling latency is now impacted both for the small- and large-scale swarm, although the effects are more pronounced in the larger system. The added latency in the centralized controller is due to drones where tasks have been assigned becoming unreachable, and the system having to reschedule and restart these tasks in new devices. The distributed framework cannot react to loss of connectivity in the same way, as decisions are made by the edge device itself. This translates to a major penalty in task execution time, which is on average 56% longer for successfully completed tasks than in the centralized framework, with an additional 18% of tasks not completing at all.

### 3.3 Realizing the Potential of Centralized Coordination Control

The results above show that both in small and large swarms centralized control outperforms distributed control in decision quality, especially in realistic swarms that are heterogeneous and unreliable. However for centralized control to scale several system challenges must be addressed. Here we show the potential of centralized control as the effect of different system bottlenecks is alleviated. From our tracing system, on average 34% of execution time is spent exchanging messages between the cloud and edge devices. The remaining 66% of time in the scheduler is spent polling the swarm state and scheduling tasks. Fig. 5 shows the scheduling latency impact from progressively lowering the network request latency, and implementing a multi-agent shared-state scheduler [19] in the centralized controller to improve scheduling throughput. Although this experiment is only a first-order approximation, it suggests that centralized control can become viable for large swarms. Techniques often found in cloud systems, such as straggler mitigation, and duplicate or hedged requests [10] can additionally optimize task execution time under a centralized control framework.

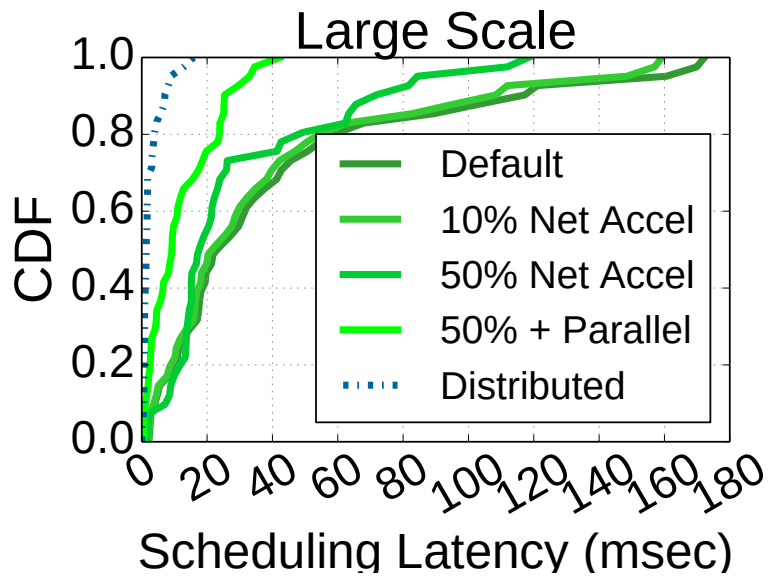

Figure 5: Scheduling latency as we progressively accelerate network processing and task scheduling.

## 4 Future Work

Swarms of autonomous edge devices are increasing in size, number, and importance. Coordination control in swarms shares a lot of commonalities with cluster management in the cloud. The goal of this work is to start the discussion on the tradeoffs between centralized and distributed coordination control of large-scale programmable swarms given the advances in low-latency network fabrics, serverless compute, and data mining frameworks. We first explore these tradeoffs in a local swarm of programmable Parrot drones, and second in swarms of thousand of heterogeneous edge devices, through simulation. We conclude that although superior in scheduling quality, centralized control needs to overcome several several challenges to scale, including: *low-latency RPCs*, *fast scheduling of very short tasks*, and *straggler mitigation techniques*.